\documentclass[prd,twocolumn,superscriptaddress,groupedaddress,amsmath,amsfonts,amssymb,showpacs,nofootinbib]{revtex4}
\pdfoutput=1
\usepackage{graphicx} 
\usepackage{dcolumn}  
\usepackage{bm}
\usepackage[usenames,dvipsnames]{color}
\usepackage[colorlinks=true, linkcolor=BrickRed, citecolor=Blue, urlcolor=Blue, filecolor=Blue]{hyperref}

\def\be{\begin{equation}}
\def\ee{\end{equation}}
\def\bea{\begin{eqnarray}}
\def\eea{\end{eqnarray}}

\def\teq{t_{\rm eq}}
\def\zeq{z_{\rm eq}}
\def\xeq{x_{\rm eq}}

\def\xi{x_{\rm i}}
\def\zi{z_{\rm i}}
\def\Ri{R_{\rm i}}
\def\ti{t_{\rm i}}

\def\xd{x_{\rm d}}
\def\td{t_{\rm d}}

\def\xc{x_{\rm c}}
\def\zc{z_{\rm c}}

\def\xf{x_{\rm f}}

\def\xto{x_{\rm to}}
\def\zto{z_{\rm to}}

\def\xta{x_{\rm TA}}
\def\rta{r_{\rm TA}}

\def\Rdec{R_{\rm dec}}

\def\Mloop{M_\mathrm{loop}}

\begin{document}

\title{Constraints on cosmic strings from ultracompact minihalos}

\author{Madeleine Anthonisen}
\affiliation{Department of Physics, McGill University, Montr\'eal, QC, H3A 2T8, Canada}

\author{Robert Brandenberger}
\affiliation{Department of Physics, McGill University, Montr\'eal, QC, H3A 2T8, Canada}

\author{Pat Scott}
\affiliation{Department of Physics, Imperial College London, Blackett Laboratory, Prince Consort Road, London SW7 2AZ, UK}


\date{\today}

\begin{abstract}

Cosmic strings are expected to form loops.  These can act as seeds for accretion of dark matter, leading to the formation of ultracompact minihalos (UCMHs).  We perform a detailed study of the accretion of dark matter onto cosmic string loops and compute the resulting mass distribution of UCMHs.  We then apply observational limits on the present-day abundance of UCMHs to derive corresponding limits on the cosmic string tension $G\mu$.  The bounds are strongly dependent upon the assumed distribution of loop velocities and their impacts on UCMH formation.  Under the assumption that a loop can move up to a thousand times its own radius and still form a UCMH, we find a limit of $G\mu\le 1\times10^{-7}$.  We show, in opposition to previous results, that strong limits on the cosmic string tension are not obtainable from UCMHs when more stringent (and realistic) requirements are placed on loop velocities.

\end{abstract}

\maketitle

\section{Introduction}
\label{intro}

Cosmic strings (see e.g. \cite{Vilenkin85, VilenkinShellard, Hindmarsh95, rhb} for reviews) are topological defects present in many theories of particle physics beyond the Standard Model. They are lines of confined energy density, analogous to defects such as vortex lines in condensed matter systems like superconductors and superfluids.  In all particle theories that permit cosmic strings, a network of strings forms during a phase transition in the very early Universe.  Causality arguments \cite{Kibble80,Kibble82} show that this network persists to the present time.  Using cosmological observations to hunt for gravitational effects of the energy trapped in cosmic strings is therefore a powerful way to probe particle physics beyond the Standard Model \cite{Brandenberger14}.

Observable signatures of cosmic strings are typically proportional to the mass per unit length $\mu$ of the string \cite{Vilenkin85, VilenkinShellard, Hindmarsh95, rhb}, which is in turn related to the energy scale $\eta$ at which the strings form ($\mu \simeq \eta^2$ \cite{Nielsen73}).  Searching for cosmological signatures of strings thus probes particle physics in a ``top down'' manner, excluding higher energy scales most easily.  This makes searches for cosmic strings highly complementary to terrestrial accelerator experiments, which search for new physics via a ``bottom up'' strategy.

As topological defects, by definition cosmic strings cannot have ends.  They must either exist as part of a network of infinite strings, or as closed loops.  Because they are relativistic, a segment of infinite string will typically have a translational velocity of the order of the speed of light.  The network of infinite cosmic strings follows a ``scaling solution" wherein the correlation length, which describes the mean curvature and separation of string segments, grows linearly with time.  This causes the contribution of the network to the energy density of the Universe to remain constant.  The analytical arguments for the existence of this solution (see e.g.\ \cite{rhb} for a review) have been confirmed by numerical simulations \cite{Albrecht85,Bennett88,Allen90,Martins06,Ringeval07,Vanchurin06,BP11}.  The scaling of the long string network is maintained by the formation of loops when segments intersect, removing energy from the network.  Studying the distribution of string loops numerically is much more demanding than following the network, because a much larger hierarchy of scales needs to be followed. However, current results indicate that loops also follow a scaling solution (see recent references in \cite{Albrecht85,Bennett88,Allen90,Martins06,Ringeval07,Vanchurin06,BP11} for numerical evidence, and \cite{Pol} for some more recent analytical work). 

String loops act as seeds for the growth of density perturbations in the matter surrounding them.  If they are present early enough, and persist for long enough, they can lead \cite{Berezinsky11} to the formation of so-called ultracompact minihalos of dark matter (UCMHs; \cite{Berezinsky03,Berezinsky06,Berezinsky07,Berezinsky08,Ricotti09,SS09,Bringmann11,Berezinsky12,Berezinsky13}).  UCMHs are distinguished from regular dark matter (DM) halos by the epoch at which they undergo gravitational collapse. Regular halos do not transition to the nonlinear regime of structure formation until $z\lesssim30$, whereas UCMHs collapse shortly after matter-radiation equality, in isolation.  At this time, the background density field is still cold, smooth and essentially featureless.  This means that UCMHs form by almost pure radial infall \cite{Ricotti09}, giving them a far steeper central density profiles than regular cold dark matter halos: $\rho\varpropto r^{-9/4}$ \cite{Fillmore84,Bertschinger85,Vogelsberger09} rather than $\rho\varpropto r^{-1}$ \cite{NFW}.  

If DM can self-annihilate, the rate goes as the square of the particle density.  The steep density profile of UCMHs therefore makes them excellent candidates for indirect detection of DM \cite{SS09,Lacki10,JG10,Berezinsky10a,Berezinsky10b,Yang12}.  Searches for gamma-ray sources with the LAT instrument aboard the Fermi gamma-ray space telescope lead to the strongest limits on the cosmological density of UCMHs \cite{Bringmann11}.  Gravitational lensing \cite{Ricotti09,Li12,Zackrisson12}, neutrinos \cite{Yang13c,Zheng14}, reionization \cite{Zhang11,Yang11a} and diffuse photon fluxes at various wavelengths \cite{Bringmann11,Yang11c,Yang13a} provide supporting limits.  With these limits and a proper understanding of how to calculate the UCMH yield from a particular scenario, it becomes possible to use UCMHs to place limits on the spectrum of primordial perturbations \cite{JG10,Bringmann11,Yang13b,Yang13c}, non-Gaussianities \cite{Shandera12} and cosmic strings \cite{Berezinsky11}.  Here we provide an improved treatment of UCMH formation around cosmic string loops, and the resulting limits on the string tension $G\mu$.

In Section \ref{scaling} we review the important aspects of the cosmic string loop scaling solution, before following the accretion of DM by loops in Section \ref{accretion},  and the subsequent formation of UCMHs.  In Section \ref{fraction} we apply those calculations to determine the fraction of DM in loop-induced UCMHs, including the effects of loop velocities.  We derive limits on the cosmic string tension in Section \ref{limits}, then summarize in Section \ref{conclusion}.  For the most part we use natural units, with $c$ set to $1$.

\section{String loop scaling}
\label{scaling}

An important ingredient in investigating the formation and evolution of dense 
structures seeded by cosmic string loops is the number density of loops per 
unit loop radius, $n(R,t)$.  The quantity $n(R,t)dR$ is the number density 
of loops in the cosmic string network at a time $t$ with radii between $R$ and 
$dR$.

In this paper we use a one-scale model for the distribution of
string loops \cite{Vilenkin81,Kibble85}, according to which all loops of initial radius $\Ri$ form
at the same time $\ti(\Ri)$.  Here $\Ri$ and $\ti(\Ri)$ are
linearly related by a constant $\alpha$:
\begin{equation}
\frac{\Ri}{t_{\rm i}(\Ri)} = \frac{\alpha}{\beta} \, .
\label{csrad}
\end{equation}
The numerical value of $\alpha$ must be determined from simulations; we adopt $\alpha=0.05$ \cite{BP11}.
String loops warp and twist as they evolve, so the probability that any given loop is exactly circular at any point in time is virtually nil.  It is common to introduce a parameter $\beta \equiv l/R$ that relates the radius of a loop to its length $l$.  Deviations from circularity can then be accounted for by allowing $\beta$ to differ slightly from $2\pi$ (although for our final limits we simply set $\beta=2\pi$).

The scaling solution implies that a constant number $N$ of loops are formed per expansion time per Hubble volume, meaning that the number density of loops at the time of their formation is 
\begin{equation}
n(\Ri,\ti) = N\ti^{-4} = N \alpha^4\beta^{-4}\Ri^{-4}.
\label{cm}
\end{equation}
$N$ is another constant that must be determined by simulations; we take $N=40$ \cite{BP11}.
 
Neglecting, for the moment, slow decay of loops by emission of gravitational radiation,
the physical radius $R$ of a string loop remains constant as the Universe expands. The number 
density redshifts, so that for $t > \ti(R)$ the number density of loops of radius $R$ is  
\begin{equation}
n(R,t) = \left[\frac{z(t)+1}{\zi(R)+1}\right]^{3}n(R,t_{\rm i}) \, ,
\label{rsnum}
\end{equation}
where $z(t)$ is the cosmological redshift, and $\zi(R) \equiv z[\ti(R)]$ is the redshift at the time that loops of radius $R$ were created. Making use of the fact that $t\varpropto (1+z)^{-3/2}$ during matter domination and $t\varpropto (1+z)^{-2}$ during radiation domination, the number density of loops is 
\begin{equation}
n(R,t) = N\alpha^2\beta^{-2}t^{-2}R^{-2},
\label{mdnd}
\end{equation}
for loops formed during matter domination, and 
\begin{equation}
n(R,t) = N\alpha^{5/2}\beta^{-5/2}t_{\rm eq}^{1/2}t^{-2}R^{-5/2}
\label{rdnd}
\end{equation}
for loops formed during radiation domination (as evaluated at some time $t>\teq$, where $\teq$ is the time of equal matter and radiation).

Emission of gravitational radiation by loops leads to a reduction of $R$ with time, at an approximately constant rate proportional to $G \mu$.  The actual rate has some distribution over a population of loops, according to the loops' individual geometries and the corresponding rate at which they can each emit gravitational radiation -- but there is a typical value associated with the typical loop oscillation frequency.  Combined with the fact that smaller loops are formed earlier, this means that there exists a radius $\Rdec$ below which loops will have typically decayed by a given time $t$,
\begin{equation}
\label{Rdecay}
\Rdec(t) = \frac{G\mu\gamma}{\beta} t,
\end{equation}
where $\gamma$ is another numerical constant determined from simulations \cite{Vachaspati85}. We will use $\gamma=10\pi$.

From this point onwards we make the ``fast decay approximation'', where we assume that loops decay essentially instantaneously.  In this approximation $R(t)=\Ri$ from the time of loop creation right up to decay, so $R$ and $\Ri$ become essentially interchangeable in all the equations that we have written so far.  Under the fast decay approximation, the impact of gravitational radiation can be neglected for all radii $R>\Rdec$.  Below this value, due to the fact that there is always a tail in the distribution of oscillation frequencies, and therefore some loops that shrink in radius at every rate less than the typical one, the physical number density of loops remaining will be proportional to $R$.  The density per unit loop radius therefore becomes independent of $R$ \cite{Vilenkin85,VilenkinShellard},
\be
\label{decayednd}
 n(R,t) = n(\Rdec,t) \qquad \mathrm{for\ } R < \Rdec,  
\ee
regardless of whether the loop was formed during matter or radiation domination.

\section{Accretion of dark matter by string loops}
\label{accretion}

Accretion of cold dark matter by a cosmic string loop leads to a spiky DM distribution \cite{Brandenberger87a, Brandenberger87b} (see also
\cite{Fillmore84, Bertschinger85}). If the resulting structure undergoes gravitational collapse sufficiently early, a UCMH will result. To determine the total fraction
of DM contained in loop-induced UCMHs at the present day, we must study the accretion of DM by string loops in some detail.

There are several pitfalls to navigate in doing this. It is not valid to simply naively take all loops at $\teq$,
and apply the growth factor from linear perturbation theory to the initial mass of each loop, as some loops may decay before such a period of accretion could become effective.  It would be equally incorrect to just eliminate loops that decay before a certain time (e.g.\ the present day, or $\teq$), as such loops may have accreted enough mass before their decay to have already created nonlinear structures -- possibly even during radiation domination.  Even if such loops do not induce nonlinear structures by the time of their decay, the velocity perturbation that they induce will persist, and continue to grow in time after $\teq$ even if the loop has already decayed.  That growth may eventually lead to gravitational collapse in time to create a UCMH, even though the loop decayed much earlier.  Thus, to study formation of UCMHs from string loops we must perform a careful study of the accretion of DM onto string loops both before and after $\teq$, and before and after their time of decay
\be
\label{td}
 \td=\frac{\alpha}{\gamma G \mu} \ti.
\ee

To compute the accretion of mass by loops of radius $R$ produced at time $\ti(R)$, we use the Zel'dovich approximation \cite{Zel70}; specifically, the spherical collapse model. We consider fluctuations that are initially isothermal, where the initial fluctuation is exclusively determined by the cosmic string loop.  The loop is the source of gravitational attraction, and over time will lead to an inhomogeneous DM distribution.
 
We focus on a spherical shell of matter of physical radius $r$ about the center of the loop.
We are interested in the time evolution of this radius, as the shell moves towards the center of the loop. It is convenient to parametrize
this radius in the form
\be \label{param}
r(x) = a(x)b(x)\zeta_i \, ,
\ee
where $a(x)$ is the scale factor, $\zeta_i$ is the comoving coordinate associated with the 
spherical shell at the initial time $t_{\rm i}$, and $b(x)$ measures the difference between the motion of the 
spherical region in the presence of the string loop compared to how it would
evolve under simple cosmological expansion. Here a convenient parameter for time is 
\be
\label{x}
x\equiv\frac{a(t)}{a(\teq)} = \frac{\zeq + 1}{z(t) + 1}.
\ee
In the case of cold dark matter we
can neglect thermal motion of matter, and in this approximation each shell
will be characterized by the relative mass fluctuation parameter 
$\Phi(r) \, \equiv \, \delta M / M(r)$, where $M(r)$ is the total mass within the shell
of initial radius $r$, and $\delta M$ is the mass fluctuation due to the string loop
(which is independent of $r$). 

As derived in \cite{Tkachev}, in the Zel'dovich approximation we have the
following equation of motion for $b(x)$:
\begin{equation}
x(x+1)\frac{d^{2}b}{dx^{2}} + \left(1+\frac{3}{2}x\right)\frac{db}{dx} + 
\frac{1}{2}\left(\frac{1+\Phi}{b^{2}}-b\right) = 0 \, .
\label{Mes}
\end{equation}
where the information about which shell we are considering is hidden in
the value of $\Phi$ for that shell.

The solutions of this equation depend on whether we are in the radiation-dominated
phase $t < \teq$ or in the matter-dominated phase $t > \teq$, and whether the loop has
decayed or not. In the radiation-dominated phase the DM is a sub-dominant 
component of matter, so on the sub-Hubble scales that we are 
considering when we study accretion by string loops, we expect only
logarithmic growth of the fluctuations after loop decay.  In the matter-dominated phase, we expect linear growth in $x$. Hence, we must 
study the solutions of Eq.~(\ref{Mes}) in various cases. We will in fact find that
the nonlinear mass grows linearly in $x$ even in the radiation phase, as
long as the seed loop is still present.

The size at time $x$ of a compact object formed via accretion about the string
loop is given by the radius of the shell which is ``turning around'' at the
time $x$. The turnaround time $t_{\rm TA}$ for a fixed shell is the time
when ${\dot{r}} = 0$. At that time, the shell of matter disconnects from the 
Hubble flow to collapse, forming a virialized clump.  Adopting the coordinates of 
Eq.~(\ref{param}), the turnaround condition becomes
\begin{equation}
b + x \frac{db}{dx} = 0 \, .
\label{ta}
\end{equation}

There is a critical turnaround time (with an associated critical redshift,  
$\zc$) after which a collapsing overdensity will not contain sufficiently pristine material to form by radial infall, and so cannot form a UCMH.  
The precise time at which the radial infall approximation breaks down and a collapsing halo can no longer be said to form a UCMH is still rather uncertain \cite{Bringmann11,Li12,Shandera12}.  Certainly a `latest collapse redshift' of $\zc\sim1000$ is a conservative and very safe choice \cite{Ricotti09}, but UCMH formation down to redshifts as low as $\zc\sim\mathcal{O}(100)$ is not inconceivable \cite{Bringmann11,Shandera12}.  Here we will assume $\zc=1000$ for our final limits on the cosmic string tension $G\mu$.  Adopting a smaller yet still plausible redshift would lead to improved constraints, because later collapse redshifts allow progressively smaller perturbations time to collapse and form UCMHs.

We must verify that $t_{\rm TA}$ for the innermost shell of DM around a loop occurs 
\textit{before} this critical collapse time, if the loop is to be said to have seeded a UCMH.  The turnaround time is affected by 
the loop decay time; the earlier a loop evaporates, the longer it will take for the turnaround to occur.

\subsection{Formation and accretion before $\teq$\\($\xi<1$, $x<1$)}
\label{beforeteq}

In this section we study the accretion of DM by a
string loop during radiation domination. The analysis of this section is applicable
to all loops formed before $\teq$, but different parts of the analysis apply depending on whether the loop decays before or after $\teq$.  We carry out the calculation in two different regimes.  The first regime covers the period 
from loop formation ($\xi$) up to either loop decay ($\xd$) or equality ($\xeq=1$), whichever is earlier.  This treatment is all that is required during radiation domination for loops that decay after equality, as in that case it is valid right up to $\teq$.  For loops that decay before equality ($\xd < 1$), the solution for the first regime must be matched on to the solution for the second at $x=\xd$.  The second regime extends from loop decay to equality $(x=1)$.  In the fast decay approximation, $\Phi = 0$ in this regime, as the loop is absent.
\subsubsection{Regime I ($x<\xd$)}
\label{regime1} 

It is well known from the study of accretion of cold dark matter onto
point seeds \cite{Brandenberger87a,Brandenberger87b} that the innermost shells are the first to
decouple from the Hubble flow, turn around and collapse back onto the seed mass. To see whether any mass shell turns around by a particular
time for a string loop with initial radius $R$, we must hence focus
on the shell of initial radius $R$. We first express the matter overdensity 
inside this shell in terms of cosmic string parameters. As long as the
loop decay can be neglected, the value of the overdensity parameter
$\Phi(R)$ is constant. Starting from
\begin{equation}
 \label{phi}
 \Phi(R) = \frac{\delta M}{M(R)},
\end{equation}
where $\delta M \equiv \Mloop = \mu \beta R$ is the mass of the loop and 
\be
M(R) = \frac{4}{3}\pi R^{3}\rho_{\rm DM}(t_{\rm i})
\ee
is the total mass of DM contained within the region of initial
radius $R$ (the loop radius) at the time of loop formation, we find
\begin{equation}
 \label{phics}
 \Phi(R) \equiv \Phi(\xi) = 4 \beta^3 \alpha^{-2} (G\mu) f_\chi^{-1} \kappa^{-1} x_i^{-1}.
\end{equation}
Here $f_\chi \equiv \Omega_{\rm DM} / \Omega_{\rm m}$ and we have used the value of $x$ at the time of loop formation
to label the innermost shell instead of the loop radius $R$.
Also, we have made use of the background Friedmann equation of motion
(after rescaling the DM density to the time $\teq$ and using
the fact that the total density at that time is twice the matter density), and defined the constant $\kappa$ by
\be
 H_{eq}^2 \, \equiv \, \kappa \teq^{-2} \, ,
\ee
i.e.
\be
 \kappa \equiv \frac {2^4 \pi G \rho_{\rm DM}(\teq) \teq^2}{3f_\chi},
\ee
which falls in the range $1/4 < \kappa < 4/9$.
 
During radiation domination $x$ is small, so the second 
derivative term in Eq.~(\ref{Mes}) becomes negligible.  With initial 
conditions $b(\xi)=1$ for $\xi \ll 1$, the approximate solution 
of (\ref{Mes}) is 
\be
 b(x)^3 = 1 - \frac{3}{2} x \Phi(R) \, ,
 \label{bphi}
\ee
for $x \gg x_i$. Inserting the result from Eq.~(\ref{phics}) into this, 
we obtain
\be
 \label{regime1_solution}
 b(x)^3 = 1 - 6 \beta^{3} \alpha^{-2} f_\chi^{-1} \kappa^{-1} (G \mu) \frac{x}{x_i}.
\ee
Inserting this result in the turnaround equation  Eq.~(\ref{ta}) yields
the following expression for the turnaround time
\bea
 \label{turnaround}
 \xta   &=& \frac{1}{2\Phi(R)} \\
  &=& 2^{-3} \beta^{-3} \alpha^2 f_\chi \kappa x_i (G \mu)^{-1} \nonumber
\eea
for the innermost shell.

Consider now shells outside the innermost one, with
initial physical radius $r > R$. For those shells the value of $\Phi$ is reduced to
\be \label{Phir}
 \Phi(r) = \left( \frac{R}{r} \right)^3 \Phi(R) \,.
\ee
The mass that has turned around by `time' $x$ is
\be
M(x) = \frac{4 \pi}{3} \rho_{\rm{DM}}(t_{\rm i}) \rta(x)^3 \, ,
\ee
where $\rta(x)$ is the initial radius of the shell turning around at time $x$.
From the first line of (\ref{turnaround}) it follows that 
\be
 x = \frac{1}{2\Phi(\rta)},
\ee
so that
\be
 \rta(x)^3 = 2xR^3\Phi(R). 
\ee
From (\ref{Phir}) the nonlinear mass that has accreted
around the loop at time $x$ is
\be
 \label{regime1_mass}
 M(x) = \frac{x}{\xta} M(R) = 2x \Mloop,
\ee
where $\xta$ still refers exclusively to the turnaround of the \textit{innermost} shell.

It is interesting to note the linear growth in $x$, the same growth
obtained in linear perturbation theory after matter-radiation equality.
It is also interesting to note that if the loop survives
until $x = 1$, the nonlinear mass which has collapsed at that
point is exactly twice the loop mass.

Now we consider under which conditions the resulting halo will actually collapse in this regime, as all halos that collapse already during radiation domination are sure to lead to UCMHs.  Collapse of the innermost shell in Regime I requires that $\xta < \mathrm{min}(\xd,1)$.  First of all, we see that only loops formed at $\ti < \alpha^{-1} \gamma G \mu \teq$ decay before equality, which corresponds to radii
\be 
 \label{smallR}
 \frac{R}{\teq} < \beta^{-1} \gamma G \mu.
\ee

First we deal with the case where decay occurs before equality ($\xd < 1$).  Noting that during radiation domination $\td/\ti = (\xd/\xi)^2$, from Eq.\ (\ref{td})
we obtain
\be
 \label{decayvalue}
 \xd = \alpha^{1/2} (\gamma G \mu)^{-1/2} \xi.
\ee
Comparing the expressions (\ref{turnaround}) and (\ref{decayvalue}), we see that if decay occurs before equality, the condition for turnaround in Regime I is independent of the
value of $x_i$ (and therefore also $R$).  For such loops, collapse in Regime I can only occur if
\be
 \label{regime1_condition_xdlt1}
 G \mu > 2^{-6} \beta^{-6} \alpha^3 \gamma \kappa^2 f_\chi^2.
\ee

For loops that decay after equality ($\xd > 1$), the condition for collapse in Regime I instead becomes 
\be
 \label{regime1_condition_xdgt1}
 G \mu > 2^{-3} \beta^{-5/2} \alpha^{3/2} \kappa f_\chi \left(\frac{R}{\teq}\right)^{1/2}.
\ee

For values of $G \mu$ larger than these critical values, nonlinear dark matter clumps will have formed before matter-radiation equality about cosmic string loops, and have therefore formed UCMHs.  For $R$ corresponding to decay before equality and values of $G \mu$ smaller than Eq. (\ref{regime1_condition_xdlt1}), we must also study how DM accretion continues between loop decay and equality to determine if UCMHs might still be created during radiation domination, but in Regime II instead of Regime I.
 
\subsubsection{Regime II ($x>\xd$)}
\label{regime2}

In cases where the loop decays before equality, we must continue to evaluate the evolution of the clump 
in the regime $\xd<x<1$.  Once again, the small $x$ approximation is valid and we can neglect the second derivative term in Eq.\ (\ref{Mes}).  The difference here with the calculation in Regime I is that we set $\Phi=0$, as the loop is absent, and use $b(x)$ from Regime I (Eq.\ \ref{bphi}) evaluated at $x=\xd$ as the initial condition.  This leads to solution
\begin{align}
 \label{regime2_solution}
 b^3(x) &= \frac{2b^3(\xd)-3x+3b^3(\xd)x + 3\xd}{2+3\xd} \nonumber \\
        &= 1 - (3/2)\xd\Phi(R)(2 + 3x)(2+3\xd)^{-1}.
\end{align}

Inserting this into the turnaround condition (Eq.~\ref{ta}) gives the turnaround time
\be
\label{regime2_xta}
\xta = \frac{1}{2\Phi(R)} + \frac{1}{3\xd\Phi(R)} - \frac12 \\
\ee
for the innermost shell.  Making the correction $\frac12\rightarrow\frac23$ to the final constant in order to match solutions exactly between Regimes I and II at $\xd$,\footnote{This is needed in order to account for the terms we neglected in Regime I when we took the approximation $\xi\ll1$.} we see that only for $G\mu$ larger than the critical value Eq.\ (\ref{regime1_condition_xdlt1}) do the second two terms give a positive correction to the result for Regime I (Eq.\ \ref{turnaround}).  This indicates (as expected) that there is always some growth in Regime II, but the correction is small, so we can see that the additional growth is minimal.  
The critical relationship between $G\mu$ and $R$ that results from demanding that collapse happens within Regime II (i.e. Eq.\ \ref{regime2_xta} $< 1$) is
\begin{align}
\label{regime2_condition}
  R/\teq < &\phantom{-}2^2\cdot3^{-2}\beta^{-1}\gamma(G\mu) \nonumber\\
           &-2^4\beta^2\alpha^{-3/2}\gamma^{1/2}f_\chi^{-1}\kappa^{-1}(G\mu)^{3/2} \nonumber\\
           &+2^{10}\cdot3^{-2}\beta^5\alpha^{-3}f_\chi^{-2}\kappa^{-2}(G\mu)^2.
\end{align}

Shells outside the innermost one also exhibit the reduced growth.  Following the same treatment as for Regime I, these will turn around at 
\be 
\label{regime2_rta}
\rta(x)^3 = 2R^3\Phi(R)\frac{2+3x}{3+2\xd^{-1}},
\ee
leading to logarithmic growth of the nonlinear mass
\be
\label{regime2_mass}
M(x) = 2\xd \Mloop\frac{3x+2}{3\xd+2}.
\ee
Here we can see both the linear growth of Regime I (the prefactor of $2\xd$), and the nonlinear growth after $\xd$ (the trailing correction).  For loops that decay before equality, the UCMH mass at equality will therefore be
\be
M(x = 1) = \frac{10\xd}{3\xd+2}\Mloop.
\ee
If $\xd > 1$ on the other hand, from Eq.\ (\ref{regime1_mass}) we see that 
\be
M(x = 1) = 2\Mloop. 
\ee

We close this subsection with a warning: one might try to argue that if a nonlinear clump forms by $x = \xd < 1$ then this clump would seed linear growth in the nonlinear mass during the period $\xd < x < 1$ in the same way that the loop itself seeds growth between $\xi < x < \xd$. This argument is incorrect because the outer shells, although sensitive to the presence of the seed loop, are not sensitive (by Birkhoff's Theorem) to a re-distribution of mass within their own radii.  This redistribution therefore leads to a nonlinear core but leaves a relative underdensity between the core radius and the shell in question.

\subsection{Formation before $\teq$, accretion after $\teq$\\($\xi<1$, $x>1$)}
\label{afterteq1}

After equality $x > 1$, so the higher-order terms in $x$ dominate the equation of motion (\ref{Mes}).  The approximate form of the equation then becomes
\begin{equation}
  \label{Meslate_initial}
  x^2\frac{d^{2}b}{dx^{2}} + \frac{3}{2}x\frac{db}{dx} + \frac{1}{2}\left(\frac{1+\Phi}{b^{2}}-b\right) = 0.
\end{equation}
Because we are only interested in $\mathcal{O}(<$1) negative corrections from unity to $b$, we can also take the term linear in $x$ and linearise it in $b$ about $b=1$.  Doing this, and recasting in terms of the comoving displacement $\Delta b(x)\equiv 1-b(x)$, the equation of motion becomes
\be
 \label{Meslate}
 2x^2 \Delta b'' + 3x \Delta b' - (2\Phi+3)\Delta b = \Phi,
\ee
where primes indicates derivatives with respect to $x$.

\subsubsection{Regime III ($\xd<1$)}
\label{regime3}

For loops that have already decayed before equality, $\Phi=0$, so the equation of motion to solve is
\be
 \label{Meslate_regime3}
 2x^2 \Delta b'' + 3x \Delta b' - 3\Delta b = 0,
\ee
which has the solution
\begin{align}
 \label{regime3_solution}
 \Delta b(x) =& \frac25x^{-3/2}\left[\Delta b(1) - \Delta b'(1)\right] \nonumber\\
              &+ \frac15x\left[3\Delta b(1) + 2\Delta b'(1)\right].
\end{align}
The initial conditions are given by $\Delta b(1)$ and $\Delta b'(1)$, as calculated during radiation domination after loop decay, i.e. in Regime II according to (\ref{regime2_solution}):
\begin{align}
\Delta b(1) &= 1 - b(1) \nonumber\\
            &= 1 - \left[1-\frac{15}{2}\xd\Phi(R)(2+3\xd)^{-1}\right]^{1/3},\\
\Delta b'(1) &= \frac32\xd\Phi(R)(2+3\xd)^{-1}b(1)^{-2}.
\end{align}

We know that $\frac32\xd\Phi(R)(2+3\xd)^{-1}<1$, so we can see that $\Delta b(1) \gg \Delta b'(1)$, allowing the $\Delta b'(1)$ terms in Eq.\ (\ref{regime3_solution}) to be neglected.  Because we are interested in $x>1$ in this regime, we can also neglect the decaying solution proportional to $x^{-3/2}$, leaving
\be
 \label{regime3_solution_approximate}
 \Delta b(x) \approx \frac35x\Delta b(1).
\ee
Inserting this into the turnaround condition (Eq.\ \ref{turnaround}) gives
\be
\label{regime3_longxta}
\frac{\xd\Phi(R)}{2+3\xd} = \frac{1}{9\xta}\left(3 - \frac{15}{6\xta} + \frac{25}{36\xta^2}\right).
\ee
Because $\xta>1$, this is approximately
\be
\xta = \frac{2+3\xd}{3\xd\Phi(R)}.
\ee
Again, this does not perfectly match onto the solutions for Regimes I and II at $x=1$ and $\xd=1$, due to the various large-$x$ approximations we have made along the way, but agreement can be forced with a simple $\mathcal{O}(1)$ correction (at the expense of accuracy at very large $x$), giving 
\be
\label{regime3_xta}
\xta = \frac{2+3\xd}{10\xd\Phi(R)}.
\ee
In terms of the radius of a shell turning around at time $x$, this gives
\be
\rta(x)^3 = 10x\xd\Phi(R)R^3(2+3\xd)^{-1},
\ee
and a UCMH mass of 
\be
 \label{regime3_mass}
 M(x) = \frac{10x\xd\Mloop}{2+3\xd}.
\ee

To have UCMH collapse occur in this regime demands $\xta < \xc$, so that with Eq.\ (\ref{regime3_xta}) the condition on $G\mu$ is
\be
 \label{regime3_condition_init}
 G\mu > 2^{-4}\cdot10^{-2}(2+3\xd)^2\xc^{-2}\beta^{-6}\alpha^3\gamma f_\chi^2\kappa^2.
\ee
This translates to
\be
 \label{regime3_condition}
 \left(\frac{R}{\teq}\right)^{1/2} < \frac23 G\mu \left[ \frac{20\xc\beta^{5/2}}{\alpha^{3/2}f_\chi\kappa} - \left(\frac{\gamma}{G\mu\beta}\right)^{1/2}\right],
\ee 
which corresponds to positive $R$ only when $G\mu > 2^{-2}\cdot10^{-2}\xc^{-2}\beta^{-6}\alpha^3\gamma f_\chi^2\kappa^2$, the value below which UCMHs cannot form in the asymptotic limit $R\to0$.

\subsubsection{Regime IV ($\xd>1$, $x<\xd$)}
\label{regime4}

For loops that have not decayed by the time of equality, calculating the accretion of matter in the period between equality and loop decay requires solving the full inhomogeneous ODE (Eq.\ \ref{Meslate}).  In this case the initial conditions at $x=1$ are given by the calculation of accretion before equality and before decay, i.e. Eq.\ (\ref{regime1_solution}) from Regime I:
\begin{align}
\Delta b(1)  &= 1 - \left(1-\frac32\Phi(R)\right)^{1/3},\\
\Delta b'(1) &= \frac{\Phi(R)}{2}b(1)^{-2}.
\end{align}
We see again that $\Delta b(1)\gg\Delta b'(1)$.

The general solution to a linear inhomogeneous ODE is the sum of the general solution to the corresponding homogeneous equation, and any specific solution to the inhomogeneous one.  Often, the simplest way to obtain the solution to an initial value problem like ours is to find the solution to the homogeneous equation that satisfies the initial conditions, and choose the specific solution to be the one obtained with trivial initial conditions $\Delta b(1) = \Delta b'(1) = 0$.  This way, the sum of the two solutions is guaranteed to satisfy the true initial conditions.  The solution to Eq.\ (\ref{Meslate}) with $\Delta b(1) = \Delta b'(1) = 0$ is 
\begin{align}
  \Delta b_{\rm I}(x) =& \frac{\Phi}{3 + 2\Phi}\left\{\frac{x^{-1/4}}{2}\left[x^{-\Psi/4}(1-\Psi^{-1})\right.\right. \nonumber\\
                       &\left.+ x^{\Psi/4}(1+\Psi^{-1})\right]-1\bigg\},
\end{align}
with $\Psi\equiv(25+16\Phi)^{1/2}$.  In general $\Phi\ll1$, so $\Psi\sim5$ and
\be
  \label{regime4_inhomogeneous_solution}
  \Delta b_{\rm I}(x) \approx \frac{\Phi}{3 + 2\Phi}\left(\frac25x^{-3/2} + \frac35x -1 \right).
\ee 

The solution to the homogeneous equation is approximately given by the solution from Regime III (Eq.\ \ref{regime3_solution_approximate}).  This is once more because $\Phi$ is small, so the $2\Phi + 3$ in Eq.\ (\ref{Meslate}) can be approximated to 3, reducing the homogeneous form of Eq.\ (\ref{Meslate}) to (\ref{Meslate_regime3}).  The decaying solution and $\Delta b'(1)$ terms can also be neglected once more, for the same reasons as in Regime III.  We can now see that because the inhomogeneous solution (Eq.\ \ref{regime4_inhomogeneous_solution}) is proportional to $\Phi$, it is much smaller than the homogeneous solution.  The homogeneous solution thus dominates entirely, and 
\begin{align}
  \Delta b(x) \approx \frac35x\Delta b(1) = \frac35x\left[1-\left(1-\frac32\Phi\right)^{1/3}\right].
\end{align}

The turnaround condition again results in a polynomial in $\xta$ similar to Eq.\ (\ref{regime3_longxta}), and the lower-order terms in $\xta$ can be neglected because $\xta>1$, giving
\be
\xta = \frac{5}{3\Phi}.
\ee 
Using this to calculate the resulting mass of UCMHs that collapse in this regime gives
\be
M(x) = \frac{3x\Mloop}{5}.
\ee
Again, we can slightly correct this expression to properly match onto the Regime I solution at $x=1$, resulting in 
\be
 \label{regime4_mass}
 M(x) = 2x\Mloop.
\ee

Using the correspondingly corrected expression for $\xta$, and demanding collapse before some final $x$ (denoted $\xf$) gives
\be
 \label{regime4_condition}
 G\mu > \frac12 \beta^{-3}\alpha^2 f_\chi\kappa \frac{\xi}{\xf}.
\ee
Here $\xf = \min(\xd,\xc)$; if decay happens before the latest allowed redshift of collapse for UCMH formation then $\xf=\xd$, otherwise $\xf=\xc$. 

\subsubsection{Regime V ($\xd>1$, $x>\xd$)}
\label{regime5}

This regime deals with the period after a decay that occurs during matter domination, but before the latest allowed redshift of UCMH collapse. To solve for the accretion history in this regime, we would take the solution at the end of Regime IV as an initial condition, and evolve further from $\xd$ using the homogeneous form of Eq.\ (\ref{Meslate}).  However, the solution we obtained for Regime IV ended up being dominated by the homogeneous solution anyway, so all calculations for Regime IV apply directly to this regime.  The only difference is that we must use $\xf=\xc$ for determining the condition for collapse to occur in this regime, even though $\xd<\xc$.  This gives
\be
 \label{regime5_condition}
 G\mu > \frac12 \beta^{-3}\alpha^2 f_\chi\kappa \frac{\xi}{\xc}.
\ee

\subsection{Formation and accretion after $\teq$\\($\xi>1$, $x>1$)}
\label{afterteq2}

For loops formed after equality, we need to solve the equation of motion in the large $x$ approximation, starting from initial conditions imposed at the time of loop formation $\xi$.

\subsubsection{Regime VI ($x<\xd$)}
\label{regime6}

Here we start from the initial conditions $b(\xi)=1$, $b'(\xi)=0$ as in Regime I (i.e. $\Delta b(\xi) = \Delta b'(\xi) = 0$), and solve the inhomogeneous equation as in Regime IV.  The solution is
\be
 \label{regime6_solution}
 \Delta b(x) = \frac{\Phi}{3+2\Phi}\left[\frac25\left(\frac{x}{\xi}\right)^{-3/2} + \frac35\left(\frac{x}{\xi}\right) - 1 \right].
\ee
The resulting turnaround condition (again discarding smaller powers of $\xta/\xi$ as $\xta\gg\xi$) is
\be
  \label{regime6_xta}
  \xta\approx\frac52\xi\left(1+\frac1\Phi\right),
\ee
which leads to the condition
\be
  \label{regime6_condition}
  G\mu > \frac{5\beta^{-3}\alpha^2 f_\chi\kappa\xi^2}{8\xf - 20\xi}
\ee
for collapse to occur in this regime.  Here $\xf$ is again the end-point of the regime, which will be the earlier of the decay time $\xd$ and the latest allowed time of UCMH collapse $\xc$.  From Eq.\ (\ref{regime6_xta}) we can also find the total mass of UCMHs collapsing in this regime,
\be
  \label{regime6_mass}
  M(x) = \left(\frac{2x}{5\xi} - 1\right)\Mloop.
\ee

\subsubsection{Regime VII ($x>\xd$)}
\label{regime7}

After loop decay, we need to continue from $\xd$ under the approximation that $\Phi=0$.  This means solving the same version of the equation of motion as in Regime III (Eq.\ \ref{Meslate_regime3}), using $\Delta b(\xd)$ and $\Delta b'(\xd)$ from Regime VI (Eq.\ \ref{regime6_solution}) as initial conditions.  This gives
\begin{align}
 \label{regime7_solution_long}
 \Delta b(x) =& \frac25\left(\frac{x}{\xd}\right)^{-3/2}\left[\Delta b(\xd) - \xd\Delta b'(\xd)\right] \nonumber\\
              &+ \frac{x}{5\xd}\left[3\Delta b(\xd) + 2\xd\Delta b'(\xd)\right].
\end{align}
As $x>\xd$ in this regime, we can drop the decaying solution once more.  We can also discard lower-order terms in $\xd/\xi$ after substituting in $\Delta b(\xd)$ and $\Delta b'(\xd)$ because $\xd > \xi$, giving
\begin{align}
 \label{regime7_solution}
 \Delta b(x) \approx \frac{3x\Phi}{5\xd(3+2\Phi)}\left(\frac\xd\xi - 1\right).
\end{align}

With this solution, turnaround takes place at
\be  
 \label{regime7_turnaround}
 \xta = \frac{5}{\xi^{-1}-\xd^{-1}}\left(\frac{1}{2\Phi}+\frac13\right).
\ee
Demanding that $\xta<\xc$ for collapse to occur in this regime gives
\be
 \label{regime7_condition_raw}
 G\mu > \frac{15\beta^{-3}\alpha^2f_\chi\kappa\xi^2\xd}{24\xc\xd - 24\xc\xi - 40\xi\xd},
\ee
with a corresponding UCMH mass of 
\be 
 \label{regime7_mass_raw}
 M(x) = \left[\frac25x(\xi^{-1} - \xd^{-1})-\frac23 \right]\Mloop.
\ee
Comparing this expression for the mass at $x=\xd$,
\be
 M(\xd) = \left[\frac25\frac\xd\xi - \frac{16}{15} \right]\Mloop,
\ee
to the corresponding expression at the end of Regime VI,
\be
 M(\xd) = \left[\frac{2}{5}\frac\xd\xi - 1 \right]\Mloop,
\ee 
we see that there is excellent agreement.  To enforce an exact match, we can simply correct Eq.\ (\ref{regime7_condition_raw}) to 
\be
 \label{regime7_condition}
 G\mu > \frac{5\beta^{-3}\alpha^2f_\chi\kappa\xi^2\xd}{8\xc\xd - 8\xc\xi - 12\xi\xd},
\ee
and Eq.\ (\ref{regime7_mass_raw}) to
\be 
 \label{regime7_mass}
 M(x) = \left[\frac25x(\xi^{-1} - \xd^{-1})-\frac{9}{15} \right]\Mloop.
\ee

\subsection{Summary} 

\begin{figure*}
  \includegraphics[scale=1, trim = 0 15 0 15, clip = true]{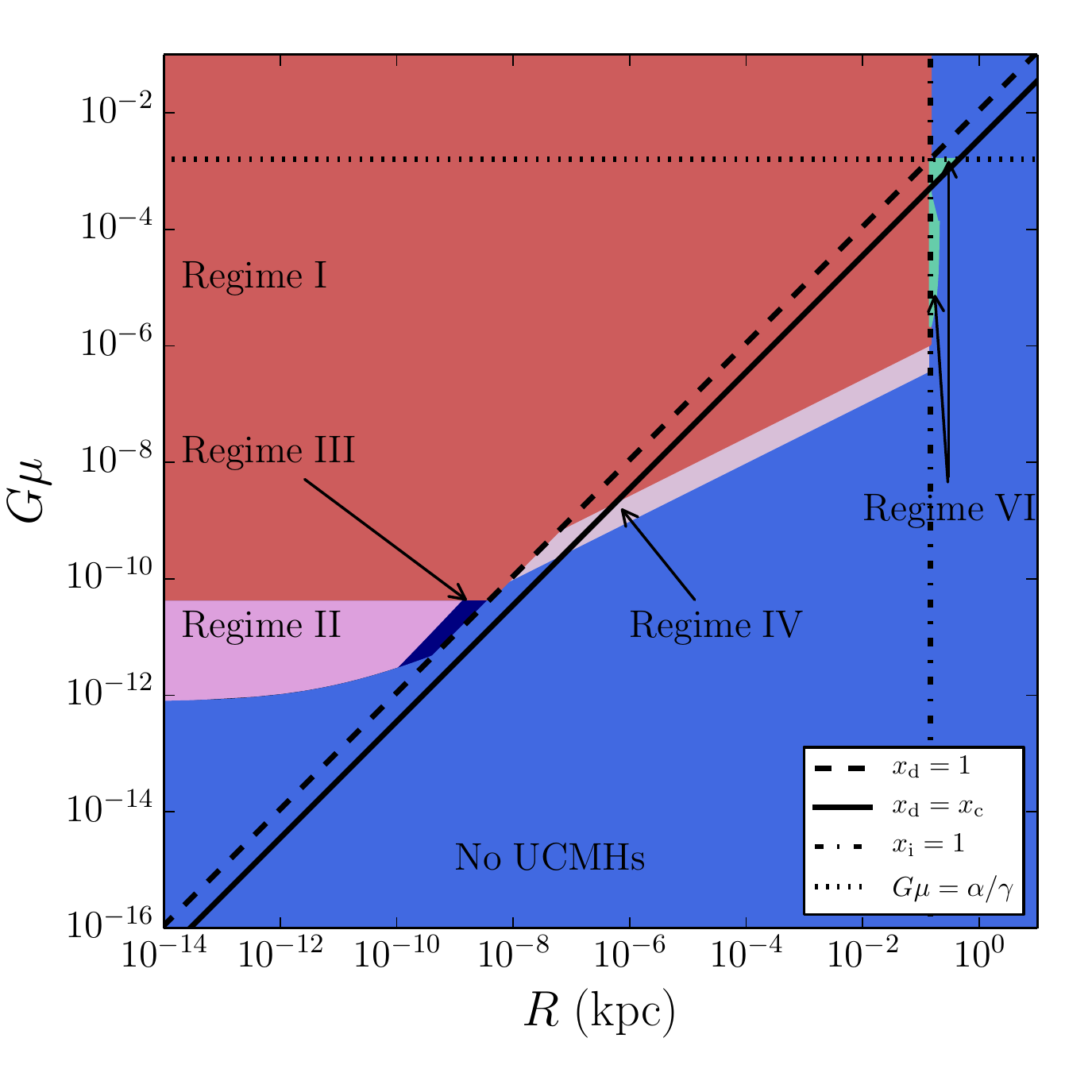}
  \caption{\label{one} Values of $G\mu$ and loop radii $R$ for which UCMHs can form.  The different regions are marked according to which regime UCMH collapse occurs in, as defined in Section \ref{accretion}.  Diagonal lines show the borders between the areas in which loop decay takes place during radiation domination, during matter domination but before the latest permitted redshift of UCMH collapse ($\zc$, which we set to 1000), and after UCMH collapse.  The vertical line shows the border between the regions where loops form during either radiation or matter domination.  The horizontal line is the largest value of $G\mu$ for which the approximation of a constant decay rate (\ref{td}), and our derived limits, hold.}
\end{figure*}

Here we summarize the salient expressions from Regimes I--VII, arranging them into a set of five different scenarios for loop formation and decay, and a scheme for determining the present-day UCMH mass $M_0$ and which (if any) regime UCMH collapse occurs in, for any given combination of $G\mu$ and $R$.  We also illustrate the resulting regions in Fig.\ \ref{one}.

The expressions delineating the different regions are given in terms of the four critical times: $\xd$, $\xi$, $\xc$ and $\xto$.  The first two follow directly from $R$,
\begin{align}
\xi &= \left(\frac{\ti}{\teq}\right)^a = \left(\frac{\beta}{\alpha}\frac{R}{\teq}\right)^a,\\
\xd &= \left(\frac{\td}{\teq}\right)^a = \left(\frac{\beta}{\gamma G\mu}\frac{R}{\teq}\right)^a,
\end{align}
where $a=1/2$ during radiation domination and $a=2/3$ during matter domination.  From these two expressions, we see that when $G\mu > \alpha/\gamma \sim 10^{-3}$, decay happens essentially immediately and UCMHs cannot form.  This constitutes a hard upper limit to the range of values of $G\mu$ that we will consider.

The second two critical times ($\xc$ and $\xto$) are independent parameters that must be chosen for a given analysis.  The latest allowed time of UCMH collapse ($\xc$) is discussed at the beginning of this Section; we adopt $\zc=1000$, which leads to $\xc=3.12$.  The `turn off' time $\xto$ is the time during late matter domination at which linear accretion onto an already-formed UCMH ceases.  As in previous papers \cite{SS09,Bringmann11}, we take this to correspond to the time at which structure formation has progressed far enough to catch most UCMHs up in bound structures and prevent further accretion from the cosmological background: $\zto\sim10$, corresponding to $\xto=284$.

\medskip
\textbf{Scenario A} ($\xi<1$ and $\xd < 1$):
\begin{itemize}\small
  \item $M_0 = 10\xto\xd\Mloop(2+3\xd)^{-1}$ (Eq.\ \ref{regime3_mass})
  \item $G\mu < $ Eq. (\ref{regime3_condition})\\$\implies$ no UCMHs
  \item Eq. (\ref{regime3_condition}) $< G\mu < $ Eq. (\ref{regime2_condition})\\$\implies$ collapse in Regime III
  \item Eq. (\ref{regime2_condition}) $< G\mu < $ Eq. (\ref{regime1_condition_xdlt1})\\$\implies$ collapse in Regime II
  \item $G\mu > $ Eq. (\ref{regime1_condition_xdlt1})\\$\implies$ collapse in Regime I
\end{itemize}

\textbf{Scenario B} ($\xi<1$ and $1 < \xd < \xc$):
\begin{itemize}\small
  \item $M_0 = 2\xto\Mloop$ (Eq.\ \ref{regime4_mass})
  \item $G\mu < $ Eq. (\ref{regime5_condition})\\$\implies$ no UCMHs
  \item Eq. (\ref{regime5_condition}) $< G\mu < $ Eq. (\ref{regime4_condition}; $\xf=\xd$)\\$\implies$ collapse in Regime V
  \item Eq. (\ref{regime4_condition}; $\xf=\xd$) $< G\mu < $ Eq. (\ref{regime1_condition_xdgt1})\\$\implies$ collapse in Regime IV
  \item $G\mu > $ Eq. (\ref{regime1_condition_xdgt1})\\$\implies$ collapse in Regime I
\end{itemize}

\textbf{Scenario C} ($\xi<1$ and $\xd > \xc$):
\begin{itemize}\small
  \item $M_0 = 2\xto\Mloop$ (Eq.\ \ref{regime4_mass})
  \item $G\mu < $ Eq. (\ref{regime4_condition}; $\xf=\xc$)\\$\implies$ no UCMHs
  \item Eq. (\ref{regime4_condition}; $\xf=\xc$) $< G\mu < $ Eq. (\ref{regime1_condition_xdgt1})\\$\implies$ collapse in Regime IV
  \item $G\mu > $ Eq. (\ref{regime1_condition_xdgt1})\\$\implies$ collapse in Regime I
\end{itemize}

\textbf{Scenario D} ($\xi>1$ and $\xd < \xc$):
\begin{itemize}\small
  \item $M_0 = \left[\frac25\xto(\xi^{-1} - \xd^{-1})-\frac{9}{15} \right]\Mloop$ (Eq.\ \ref{regime7_mass})
  \item $G\mu < $ Eq. (\ref{regime7_condition})\\$\implies$ no UCMHs
  \item Eq. (\ref{regime7_condition}) $< G\mu < $ Eq. (\ref{regime6_condition}; $\xf=\xd$)\\$\implies$ collapse in Regime VII
  \item $G\mu > $ Eq. (\ref{regime6_condition}; $\xf=\xd$)\\$\implies$ collapse in Regime VI

\end{itemize}

\textbf{Scenario E} ($\xi>1$ and $\xd > \xc$):
\begin{itemize}\small
  \item $M_0 = \left[\frac{2}{5}\xto\xi^{-1} - 1\right]\Mloop$ (Eq.\ \ref{regime6_mass})
  \item $G\mu < $ Eq. (\ref{regime6_condition}; $\xf=\xc)$\\$\implies$ no UCMHs
  \item $G\mu > $ Eq. (\ref{regime6_condition}; $\xf=\xc)$\\$\implies$ collapse in Regime VI
\end{itemize}

\section{Ultracompact minihalos from cosmic strings}
\label{fraction}

\subsection{Cosmological density of UCMHs}

The actual physical number density of loops with radius greater than or equal to $R$ at time $t$ is the integral of the number density per loop radius, over all loop radii greater than $R$,
\be
 \label{physical_number_density}
 n_{\rm phys}(R,t_0) = \int_R^{R_{\rm max}} n(R',t_0) dR'.
\ee
Here $R_{\rm max}$ refers to some maximum loop radius, e.g. the radius of loops created at equality if the number density of loops with $\xi<1$ is sought.  Eq. (\ref{physical_number_density}) gives $n_{\rm phys}(R,t) \varpropto n(R,t)R$ for $R_{\rm max}\gg R$. Referring to Eqs.\ (\ref{mdnd}) and (\ref{rdnd}), we see that $n_{\rm phys}(R,t)\varpropto R^{-1}$ for loops created during matter domination, and $n_{\rm phys}(R,t)\varpropto R^{-3/2}$ for loops born during the radiation era.  In both cases the physical density is a steeply-falling function of the loop radius.

The total present-day cosmological density of the dark matter in UCMHs, from cosmic strings of radius greater than or equal to $R$, is likewise given by 
\be
 \label{Omega_UCMH}
 \Omega_{\rm UCMH}(G\mu,R) = \rho_{\rm c}^{-1}\int_R^{R_{\rm max}} n(R',t_0) M_0(G\mu,R') dR'.
\ee
The fraction $f_{\rm UCMH}\equiv\Omega_{\rm UCMH}/\Omega_{\rm DM}$ of dark matter in UCMHs today is then simply 
\be
 \label{f_UCMH}
 f_{\rm UCMH} = \rho_{\rm DM}^{-1}\int_R^{R_{\rm max}} n(R',t_0) M_0(G\mu,R') dR'.
\ee
In contrast to the physical loop number density however, these two expressions need \textit{not} account for the reduction in the number density of loops that have decayed by $t_0$, as the calculations of the previous section ensure that the impacts of loop decay are already accounted for in the expressions for $M_0$.  Therefore, $n(R,t_0)$ here should be understood to simply be Eq.\ (\ref{mdnd}) or (\ref{rdnd}) (depending on when the loops were formed), without the correction below $\Rdec$ given in Eq.\ (\ref{Rdecay}).  For loops created during radiation domination, $M_0\varpropto R^{3/2}$ if $\xd<1$ and $M_0\varpropto R$ if $\xd>1$.  For loops created during matter domination, $M_0\varpropto R^{1/3}$.  The contribution to $f_{\rm UCMH}$ from loops decaying after equality is therefore dominated by those that decay close to equality.

\begin{figure*}
  \includegraphics[scale=0.9]{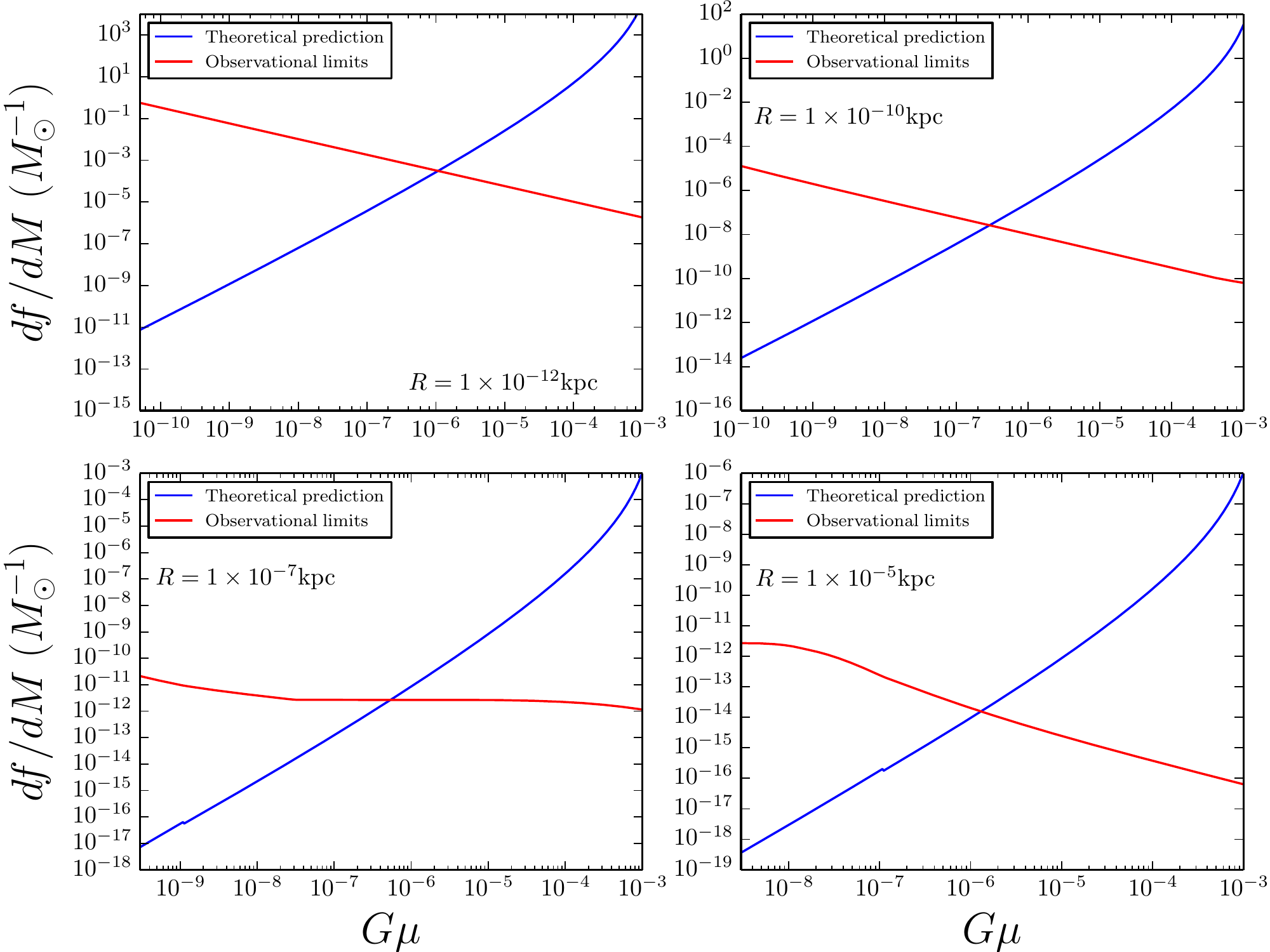}
  \caption{\label{fig2} Some example comparisons between observational limits on the UCMH abundance, and the abundance predicted on the basis of cosmic strings.   Here we show limits and predictions in terms of the differential fraction of dark matter in UCMHs as a function of $G\mu$, for four example loop radii.  For these examples, we assume that a loop can move a distance of $1000R$ during its lifetime and still form a UCMH, and adopt a $100$\,GeV DM candidate that annihilates into $b\bar{b}$ pairs with cross-section $3\times10^{-26}$\,cm$^3$\,s$^{-1}$.}
\end{figure*}

However, the available observational limits \cite{Bringmann11} on the number density of UCMHs apply to each UCMH mass independently, rather than to integrated mass ranges.  These limits have been derived assuming a single UCMH mass, and therefore effectively apply to delta-function mass spectra. They can be easily converted to \textit{differential} limits on $f_{\rm UCMH}$ simply by dividing by the UCMH mass that they apply to, giving an upper limit on $\mathrm{d}f_{\rm UCMH}/\mathrm{d}M_0$ as a function of $R$.  We show these limits in Fig.\ \ref{fig2}, along with the theoretical predictions, for different loop tensions and radii.

The differential contribution to the fraction of DM in UCMHs of mass $M_0(R,G\mu)$ for a given combination of $R$ and $G\mu$ is given by
\begin{align} \label{massfraction}
\frac{\mathrm{d}f_{\rm UCMH}}{\mathrm{d}M_0} & = \frac{\mathrm{d}f_{\rm UCMH}}{\mathrm{d}R} \Bigg/ \frac{\mathrm{d}M_0}{\mathrm{d}R} \nonumber\\
                                             & = \rho_{\rm DM}^{-1} n(R,t_0) M_0(G\mu,R) \Bigg/ \frac{\mathrm{d}M_0}{\mathrm{d}R}    \nonumber\\
                                             & = C\rho_{\rm DM}^{-1} n(R,t_0) R,
\end{align}
where, as in Eqs.\ (\ref{Omega_UCMH}) and (\ref{f_UCMH}), $n(R,t_0)$ is the number density \textit{not} corrected for loop decay.  $C$ here is an order unity correction factor, ranging from $2/3$ to $3$ depending on the values of $R$ and $G\mu$, and essentially dictated by the inverse of the scaling of $M_0$ with $R$:
\begin{align}
&C(\xi<1,\xd<1) &=& &\frac{2 + 3\xd}{3+3\xd},\\
&C(\xi<1,\xd>1) &=& &1,\\
&C(\xi>1,\xd<\xc) &=& &\frac{6\xto(\xi^{-1} - \xd^{-1})-9}{2\xto(\xi^{-1} - \xd^{-1})-9},\\
&C(\xi>1,\xd>\xc) &=& &\frac{6\xto-15\xi}{2\xto-15\xi}.
\end{align}
We see now that unlike the total cosmological density of UCMHs, the contribution to the \textit{differential} UCMH fraction has approximately the same dependence on $R$ as the physical UCMH number density, and is always dominated by smaller loops.  The strongest limits on $G\mu$ are therefore expected to come from the smallest values of $R$, as the observational limits fall off less steeply at low mass than $\mathrm{d}f_{\rm UCMH}/\mathrm{d}M_0 \varpropto M^{-3/2}$ (Fig.\ \ref{fig2}; \cite{Bringmann11}).  

\subsection{Effects of cosmic string velocities}
\label{csvelocity}

Our calculations up to this point have assumed that all cosmic string
loops are created at rest. Recent numerical simulations
(e.g. \cite{BP11}) show, however, that string loops are
typically born with translational velocities that are a significant
fraction of the speed of light. These large velocities are
induced by the relativistic velocities of the long string
segments from which the string loops are split off. After
a string loop forms, its physical velocity is redshifted. 

Accretion onto a moving loop is not spherically symmetric, and
is less efficient than accretion onto a static loop.
We take into account loop velocities in an approximate manner, using the spherical growth formulae of the previous sections
provided that the physical distance the loop has moved by the time it decays is smaller than some multiple $k$ of the loop radius.  If the loop moves further than this, we assume that no UCMH forms.  

For an initial physical velocity $v_i$, the distance a loop of
radius $R$ has moved by the time it decays is  
\be
\Delta r(R)   = a(t_{\rm d})\int_{\ti}^{\td} \left( \frac{a(\ti)}{a(t)} \right)^2 \frac{v_{\rm i}}{a(\ti)}\mathrm{d}t.
\ee

Evaluating this expression for loops that decay during radiation domination, we find that in order to have $\Delta r(R)  < KR$, 
\be 
v_{\rm i}  < K\frac{(\alpha\gamma G \mu)^{1/2}}{\beta\ln\left(\frac{\alpha}{\gamma G \mu}\right)}.
\label{velcond}
\ee
where $K$ is a constant whose value we will vary.

We take the distribution of initial velocities in each of three spatial directions to be Gaussian, as this provides a good fit to the results of numerical simulations \cite{BP11,Berezinsky11}, leading to a speed distribution
\be
P(v)=\frac{2^{1/2}v^{2}}{\pi^{1/2}\langle v^{2} \rangle^{3/2}}e^{-v^{2}/2\langle v^{2} \rangle} \,.
\label{range}
\ee
The rate of UCMH formation becomes suppressed by a multiplicative factor $\cal S$, given by the integral of (\ref{range}) from zero to $v_i$:
\be
{\cal S} = \frac{2^{1/2}v_{\rm i}^{3}} {3\pi^{1/2}\langle v^{2} \rangle^{3/2}}.
\label{Sfactor}
\ee
Here we have Taylor-expanded the result about $v_i=0$ and kept only the first term; this is a very good approximation for ${\cal O}(K)\lesssim 10^3$.  In principle the distribution (\ref{range}) should also be cutoff at the speed of light, but for sufficiently small $\langle v^{2} \rangle$ ($\lesssim 0.1$), the correction this induces is negligible.  We will use the same suppression formula for all loops, even those that decay in the matter phase; the error this induces is also negligible.  

We adopt the value $\langle v^{2} \rangle^{1/2}=0.3$, assuming that the loop velocity distribution is similar to the long-string one.  This quantity has been reported to be as low as 0.15 \cite{Berezinsky11} or as high as 0.67 \cite{BP11}.  Given this uncertainty, and the difficulty in choosing an appropriate value of $K$, we show results for a range of possible velocity suppressions.  Because $K$ and $\langle v^{2} \rangle^{1/2}$ are degenerate in their effects on ${\cal S}$, we simply vary $K$ for the purposes of this illustration, using $K = x_{\rm c}, 10$ and 1000.  For $K=1$, there is essentially no limit, as in this case the assumption of a constant loop decay rate (\ref{td}) breaks down.  We also show results with no suppression (${\cal S}=1$).

Multiplying the differential contribution (\ref{massfraction}) by the velocity suppression factor ${\cal S}$ and substituting in the loop distribution (\ref{rdnd}), we obtain
\be
 \label{dfdM1}
\frac{\mathrm{d}f_{\rm UCMH}}{\mathrm{d}M_0}={\cal S} \frac{16\pi G C N \alpha^{2}}{3R\beta^{2}f_{\chi}\kappa}X^{1/2},
\ee
where $X= \alpha t_{\rm eq}/(\beta R)$ for loops formed in the radiation dominated era and $X=1$ for loops formed during matter domination.
Inserting the values from (\ref{velcond}) and (\ref{Sfactor}) into (\ref{massfraction}), we obtain the
following final formula for the differential fraction of mass in UCMHs induced by string loops formed during radiation domination:
\bea \label{final}
\frac{\mathrm{d}f_{\rm UCMH}}{\mathrm{d}M_0} \, &=& \, \frac{16(2\pi)^{1/2}\alpha^4K^3NC}{9\kappa f_{\chi}\beta^{11/2}}\frac{G}{t_0} \left(\frac{\gamma G\mu}{\langle v^{2} \rangle}\frac{t_0}{R} \right)^{3/2} \nonumber \\
                                                & & \times \left(\zeq+1\right)^{-3/4}\ln^{-3}\left(\frac{\alpha}{\gamma G\mu}\right)\, .
\eea
This scales as $K^3$, as $R^{-3/2}$ and as $(G \mu)^{3/2}$ (modulo logarithmic corrections
in $G \mu$).

Note that our treatment of loop velocities differs in one major respect from that of \cite{Berezinsky11}: their corresponding expression for (\ref{velcond}) is missing a factor of $\sqrt{G\mu}$, which arises from a missing factor of $\sqrt{t_{\rm d}}$.  Here we account for this extra factor, drastically weakening our limits in comparison to theirs.

\section{Bounds on the cosmic string tension}
\label{limits}

The fraction of DM in UCMHs of various masses is constrained \cite{Bringmann11} by the absence of dark matter self-annihilation signals observed by the \textit{Fermi}-LAT telescope (see \cite{Atwood09, LATsat}).  In this subsection, we will apply these observational constraints on $\mathrm{d}f_{\rm UCMH}/\mathrm{d}M_0 $ to derive limits on the cosmic string tension, using the theory of loop-induced UCMH formation that we developed in Section \ref{accretion}.  

Depending on the UCMH mass, the strongest limits may come from Galactic point sources, extragalactic point sources or the contribution of UCMHs to the Galactic diffuse gamma ray emission. The limits are summarized in Fig\ \ref{fig2} (red curves), where $\mathrm{d}f_{\rm UCMH}/\mathrm{d}M_0 $ varies in terms of $G\mu$, which translates into a variation in $M_{0}$ for a fixed value of $R$.  Based on Section \ref{accretion}, we also show the predicted differential fraction of DM in UCMHs as blue curves in Fig.\ \ref{fig2}.  A constraint on $G\mu$ can thus be obtained for an associated $R$ value.  In Fig.\ \ref{fig2}, we show that for a few example radii, the limiting value of $G\mu$ is at the intersection of the two curves; values of $G\mu$ for which the blue curve exceeds the red curve are observationally ruled out. Note that the scaling with $G \mu$ agrees with what is expected from (\ref{final}).

Here we carry out the usual calculation of the UCMH core radius \cite{Bringmann11} for each combination of $R$ and $G\mu$, taking the larger of the core radii implied by annihilation and angular momentum of dark matter.  We further supplement this selection with a comparison to $R$ itself, also demanding that the core radius must be at least as large as the loop.  This, taken together with the fact that the loop radius is by definition far smaller than the actual halo turnaround radius, ensures that the non-sphericity of the loop does not interfere with the radial infall approximation.

We assume a dark matter mass of 100\,GeV, a canonical thermal relic annihilation cross-section of $3\times10^{-26}$\,cm$^3$\,s$^{-1}$ and 100\% annihilation into $b\bar b$ quark pairs.  These are relatively typical parameters for WIMP dark matter, and at about the limit of what is currently allowed by CMB and gamma-ray searches for dark matter annihilation \cite{Cline13,Planck15cosmo,FermiPass8Dwarf,Slatyer15a}. It is however worth noting that the cross-section could be lower, the mass higher, or the final states less conducive to gamma-ray production than we have assumed.  Although this would further weaken the resulting limits on $G\mu$, in general UCMH limits on cosmological scenarios are not strongly dependent upon the particle nature of dark matter (so long as it can actually annihilate) \cite{Bringmann11}.

\begin{figure}
  \includegraphics[width=\columnwidth, trim = 0 20 0 10 0, clip = true]{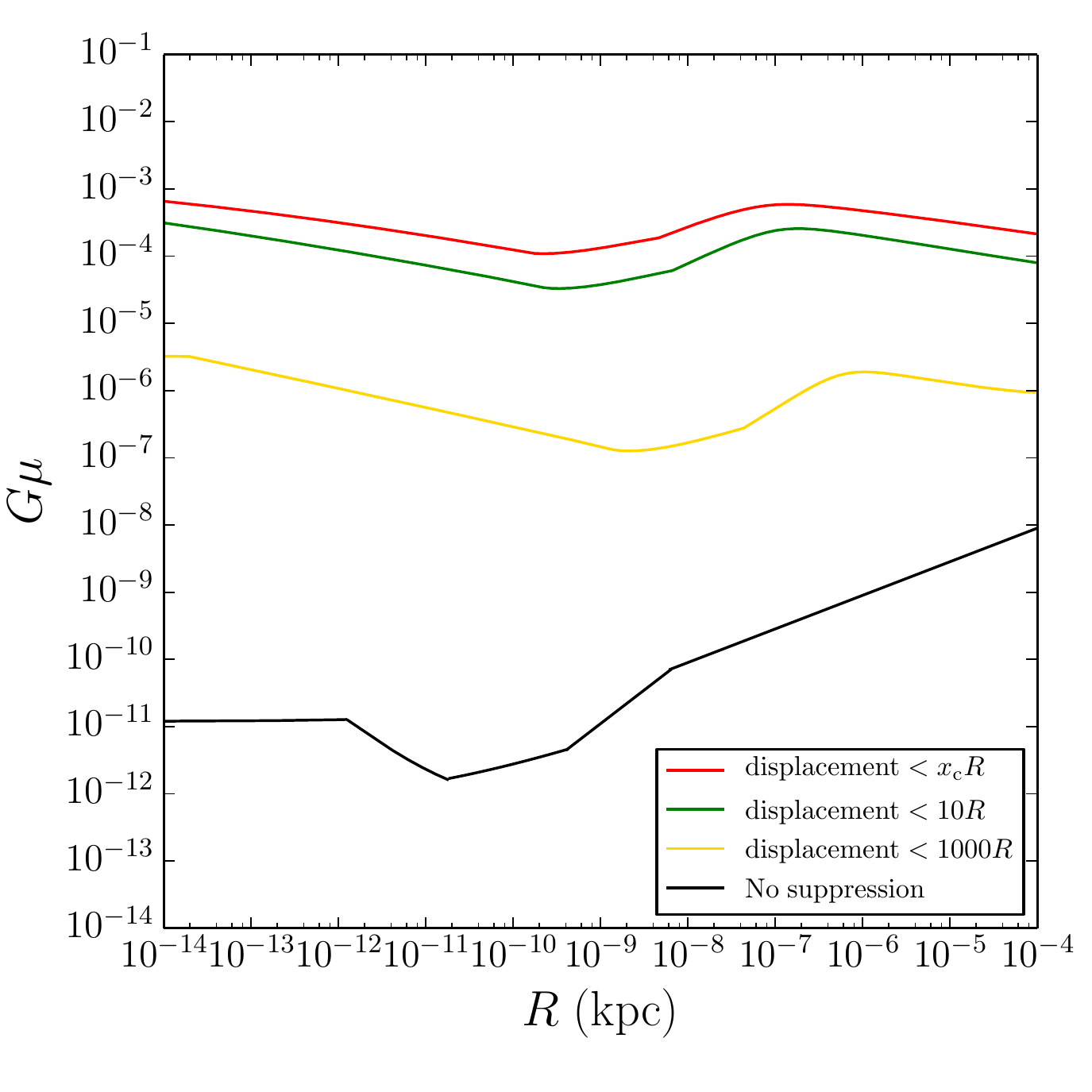}
  \caption{\label{fig3} Upper limits on $G\mu$ as a function of loop radius, for different velocity suppression factors.  The limits have little dependence on the adopted DM model; we assume a $100$\,GeV DM candidate that annihilates into $b\bar{b}$ pairs with cross-section $3\times10^{-26}$\,cm$^3$\,s$^{-1}$.}
\end{figure}

We plot our final constraints on $G\mu$ in Fig.\ \ref{fig3}, as a function of the loop radius $R$ giving rise to the limit.  It is worth remembering that because of the scaling solution, loops with all radii will be present -- so the resulting limit on $G\mu$ is simply the strongest limit available at any $R$.  The different curves we plot in this figure are based on different assumed values of the velocity suppression factor ${\cal S}$, assuming that the loop can move a different number of times its own radius before UCMH collapse is rendered impossible.

The shapes of the limit curves can mostly be understood by considering how the observational limits vary over $R$ (and correspondingly over $M_{0}$).  At the high mass (high $R$) end, extragalactic sources lead to the tightest constraints.  At intermediate masses, galactic point sources provide the best limits.  At the lowest values of $R$, it is the diffuse gamma ray emission that leads to the best constraints.  The kinks in the curves below $R\sim10^{-8}$\,kpc correspond to the transition from galactic point source limits to extragalactic sources.  Ignoring velocity suppression, we see that except at very small $R$, the shape of the limit curve tracks the theoretical minimum value of $G\mu$ at which cosmic strings can form UCMHs, rather than the observational limits on the UCMH abundance (compare to Fig.\ \ref{one}).

Clearly the ${\cal S}$-factor selected can significantly impact the constraints on $G\mu$.  The limits improve as loops are allowed to travel further during their lifetimes, as a greater fraction of loops are assumed to be able to form UCMHs.  It is only when loops are allowed to travel nearly $1000$ times their radii before UCMH collapse that the bound on $G\mu$ becomes stronger than the accepted upper limit from CMB physics of $G\mu \le 1.7 \times 10^{-7}$ \cite{Dvorkin11}.  At $K=1000$, the best constraint is $G\mu \le 1 \times 10^{-7}$ for an associated UCMH mass of $\sim 10^{2} M_\odot$.  These results are clearly inconsistent with those of \cite{Berezinsky11}, who find limits of a similar order to this when assuming just $K=1$.

Neglecting to consider the ${\cal S}$-factor entirely, equivalent to cosmic string loops either having \textit{no} translational velocity or their velocities being irrelevant to their ability to form UCMHs, improves the upper limit on $G\mu$ by several orders of magnitude.  The best constraint, at a UCMH mass of $3\times 10^{-3} M_\odot$, is $G\mu \le 2 \times 10^{-12}$.  Of course, the most physically rigorous treatment demands a velocity correction factor, so this case is not realistic.  The results show how small differences in the assumptions with regard to this aspect can propagate into order of magnitude differences in the constraint on $G\mu$.  Our results can therefore be considered to present limits ranging from ``best case" to ``worst case" scenarios.  Improved knowledge of the loop velocity distribution and simulation of UCMH formation around a moving seed would significantly increase the precision of the bounds.

The final bound on $G \mu$ is also, unfortunately, sensitive to the parameters that describe the cosmic string loop distribution, in particular the values of $\alpha$ and $\beta$. For the values we have used in this paper, drawn from \cite{BP11}, our best limit on $G \mu$ with e.g. $K=1000$ is $G\mu \le 1 \times 10^{-7}$.  However, for $\alpha=0.5$, we would obtain a more stringent bound: $G\mu \le 10^{-9}$. However, one cannot simply extrapolate the constraint obtained when increasing $\alpha$ without taking into account the corresponding increase in the lower value of $G \mu$ below which no UCMHs form (see e.g.\ \ref{regime3_condition}).  Similarly, if we reduce $\beta$ by a factor of ten, we also obtain a tighter bound: $G\mu \le 2 \times10^{-10}$.

The largest uncertainty to our results, however, comes
from the treatment of the effect of loop velocities.
We have modelled the effects with a factor of $K$ which
we varied, $K = x_{\rm c}$ being the most restrictive assumption
and larger values of $K$ being less restrictive. For this
issue, an improved analysis might be possible. For example,
we could adopt a ``delayed start approximation" in which
instead of removing the effect of loops which have too large
initial velocities, we follow the loops and let the
accretion start at a later time when the velocity has
redshifted to a sufficiently low value (we thank the Referee
for making this suggestion). Another way to improve on
our present analysis would be to study the cylindrical
accretion of matter onto a loop which is initially
rapidly moving and to use the part of that mass which
is sufficiently spherically distributed to yield an
ultra-compact object. We leave this issue for followup
work.

\section{Conclusions}
\label{conclusion}

Because many particle physics models beyond the Standard Model give rise to cosmic strings, it is interesting to explore bounds
on the cosmic string tension $\mu$ from a variety of cosmological observations. In this paper we have considered what limits can be set on the basis of the non-observation of gamma ray signals from DM annihilation events in ultra-compact mini-halos (UCMHs).

UCMHs would form by the accretion of cold dark matter by cosmic string loops at high redshift. The number of such UCMHs increases as $G \mu$ decreases. Assuming that the DM is self-annihilating, we obtain bounds on the cosmic string tension.  These bounds depend very sensitively on the parameters describing the cosmic string loop distribution, in particular the value of $\alpha$, as well as on the parameters that describe the loop velocity distribution.  Assuming that loops are still able to form UCMHs even if they move quickly enough to travel a thousand times their radii before the nascent UCMH collapses, we derive a limit of $G\mu \le 1 \times 10^{-7}$.  Tightening this assumption however leads to much weaker limits.

Our basic method is similar to that of \cite{Berezinsky11}, but our final results are not consistent with theirs, mainly due to a more careful treatment of the velocity suppression factor on our part. 

Here we have also developed the analytical theory of the accretion process in great detail, which we hope will prove a useful reference for future work on loop-induced UCMHs.  Although we have shown that this method only yields improved constraints over the currently accepted upper limit on $G\mu$ when unrealistic assumptions are made about loop velocity distributions, we have significantly improved the theoretical understanding of minihalo formation from cosmic strings, and the accuracy of gamma-ray limits on $G\mu$.  Our results also suggest that a better understanding of the velocity distribution of loops and the formation of UCMHs around moving seeds might be fruitful for helping to constrain $G\mu$.

All of the numerical UCMH routines used in this paper, as well as those from earlier papers \cite{SS09, Bringmann11, Shandera12}, can now be found in v5.1.2 of the public software package \textsf{DarkSUSY} (\href{http://www.darksusy.org}{www.darksusy.org} \cite{DarkSUSY}).

\section{Acknowledgements}
We thank Franc Duplessis and J\'{e}r\^{o}me Quintin for helpful discussions.  MA and RB are supported by an NSERC Discovery Grant to RB, and by funds from the Canada Research Chair program.  PS acknowledges funding support from the UK Science and Technology Facilities Council Ernest Rutherford scheme.

\bibliography{DMbiblio,CS}

\end{document}